**Going Out of Business: Auction House Behavior in the Massively Multi-Player Online Game**

*Glitch*

Anders Drachen, Northeastern University, 360 Huntington Avenue, Boston MA 02115; andersdrachen@gmail.com **(corresponding author, +1 857 241 8560)**

Joseph Riley, Northwestern University; 1501 Maple Avenue, Evanston IL 60201; riley.jos@gmail.com

Shawna Baskin, Northwestern University; 1331 W Fargo Apt 1301, Chicago IL 60626; shawnabaskin@gmail.com

Diego Klabjan, Industrial Engineering and Management Sciences, Northwestern University; 2145 Sheridan Road, IEMS, Evanston IL 60208; d-klabjan@northwestern.edu

**Abstract**

The in-game economies of massively multi-player online games (MMOGs) are complex systems that have to be carefully designed and managed. This paper presents the results of an analysis of auction house data from the MMOG *Glitch*, across a 14 month time period, the entire lifetime of the game. The data comprise almost 3 million data points, over 20,000 unique players and more than 650 products. Furthermore, an interactive visualization, based on Sankey flow diagrams, is presented which shows the proportion of the different clusters across each time bin, as well as the flow of players between clusters. The diagram allows evaluation of migration of players between clusters as a function of time, as well as churn analysis. The presented work provides a template analysis and visualization model for progression-based or temporal-based analysis of player behavior broadly applicable to games.

**Keywords:** virtual economy, massively multi-player online game, game analytics, auction house, longitudinal analysis

**1. Introduction**

Online games form a major component of the games industry, and have expanded strongly in terms of market share, variety and market penetration in recent years, notably due to the increasing availability of mobile platforms and the introduction of Free-to-Play (F2P) business models by the interactive entertainment industry [15,29,50,51].

Of the wide variety of online games, the Massively Multi-Player Online Game (MMOG) format, and its derivatives, is unique in that these games see thousands or more players interacting within the same virtual environment [21,22,42,46,64]. The games can support complex virtual societies that include in-game economies [3,8]. The economic systems operating inside virtual worlds, as well as the economics surrounding virtual worlds and the trade occurring between the real world and the virtual, e.g. purchasing of in-game currency using real-world currency, or vise-versa, has formed the basis for research interest from industry and academia, for the latter notably from the perspective of behavioral economics and sociology. This partly because of the sheer size of these virtual markets, the unique challenges imposed by virtual property rights [7,9,22], social and societal aspects [16] even subversive criminal activity within these worlds, notably gold farming [1,19,25]. However, perhaps more importantly because MMOGs form semi-controlled/contained environments for economic and behavioral research [24, 26,43,65]. MMOGs thus offer an environment for examining human behavior, e.g. for socioeconomic experimentation.

The work presented here is situated within the domain of economics-focused game analytics, focusing on combining dimensionality reduction of economic data with temporal pattern identification. There are two overall goals: 1) to provide a longitudinal analysis of the auction house data from the Free-to-Play, browser-based MMOG *Glitch* (Figure 1) with the purpose of uncovering patterns in the player behavior in connection with the in-game economy of the game. The analysis combines dimensionality reduction techniques with temporal analysis to enable the definition of clusters of player behavior, and how these evolve across time. 2) To develop an interactive visualization of the data, which enables non-experts to interact with the data and results from the analysis.

The motivation for the first goal is founded in previous academic work on online economies but departs from the most previous work in that it does not attempt to fit e.g. a financial theory to a MMOG market, test specific hypotheses, or examine the relationship between real-world and virtual-world economies, but rather works in a data-driven, explorative fashion focusing on the in-game economy itself, the behavior of the players in relation to it, and how to visualize these behaviors.

The motivation for the second goal is founded in the increasing need in the industry (and game science) for techniques that allow analysts to provide analysis results to a variety of stakeholders in the industry – including game designers, producers, marketing, community management etc. – in a way that allows these varied groups to dynamically interact with the data, without requiring extensive training or data mining expertise [11-14]. Data visualization is a well-established field in its own right [34], but the history of applying theories and techniques from data visualization to the specific context of game analytics is more recent [23].

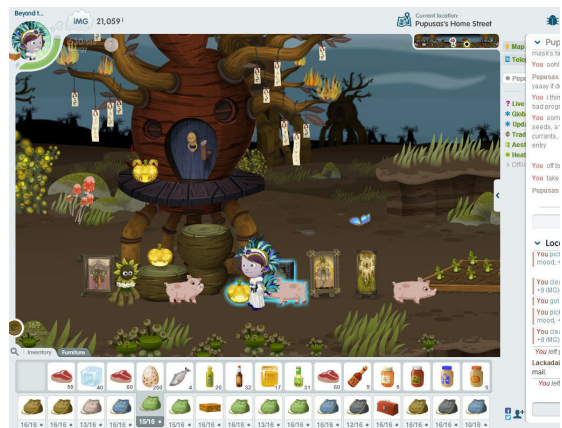

*Figure 1: Screenshot from the MMOG Glitch, showing a player's property.*

## 2. Contribution and results
The work presented consists of three parts:

1) A top-down explorative analysis of the auction house data from the MMORPG *Glitch*, describing the general features and behaviors of the dataset from the launch of the game until it was taken offline. This provides insights into the online economy of an MMORPG across its lifetime.

2) A cluster analysis of how *Glitch* players utilized – with varying degrees of success – the auction house, across a 14 month time period. The data comprise almost 3 million data points, over 20,000 unique players and more than 650 products. For the cluster analysis, 5 Key Performance Indicators (KPIs) were defined to form as the basis for the analysis. K-means clustering was applied across 14 time bins, enabling analysis of the temporal pattern of how the different clusters performed. Applying dimensionality reduction techniques on a multi-dimensional and large dataset allows for identification of patterns in player behavior [11,31]. The results show, for example, that clusters of behavior are relatively consistent in relative proportion and similarly relatively persistent, despite varying player populations in *Glitch* over the studied time frame. The four high-level clusters (Casual, Moderate, Forum, Hardcore) survive a reduction in the player base from 4,632 unique monthly players to just 723 in the final month before *Glitch* closed. The sub-types comprising each cluster do see some change, however. Retention rates fluctuate substantially across time, but apparently not as a direct effect of the *Glitch* closing down – towards the end, a 40% increase in player numbers occurred across 3 months, just 2 months before the game closed. Of the high-level clusters, only the Hardcore cluster consistently sees a large proportion of the constituent players migrating to the same behavioral cluster (Hardcore) in succeeding mounts, whereas churn rates are high for the Casual cluster.

3) A web-based interactive Sankey diagram, which shows the proportion of the different clusters across each time bin, as well as the flow of players between clusters. The diagram allows evaluation of migration of players between clusters as a function of time, as well as churn analysis. The diagram is available on: http://powerful-meadow-8588.herokuapp.com/. Sankey diagrams (Figure 2) form a type of flow diagrams, and are commonly used for analysis and visualization of energy, liquids, materials or costs of transfers between nodes or processes [27,28].

The contribution of the presented work resides both in the description of the *Glitch* economy as it evolves over time, which adds to the existing body of knowledge of virtual economies by presenting a first-time *game lifetime* view on key economic indicators; and also in the presentation and evaluation of a new method for progression-based analysis of player behavior. In combination, the analysis and Sankey diagram-based visualization combines the considerations of behavioral economic analysis with data visualization, and provides a template model for conducting progression-based or temporal-based analysis of player behavior in any digital game, beyond MMOGs, across any dimension of player behavior – not limited to economic data.

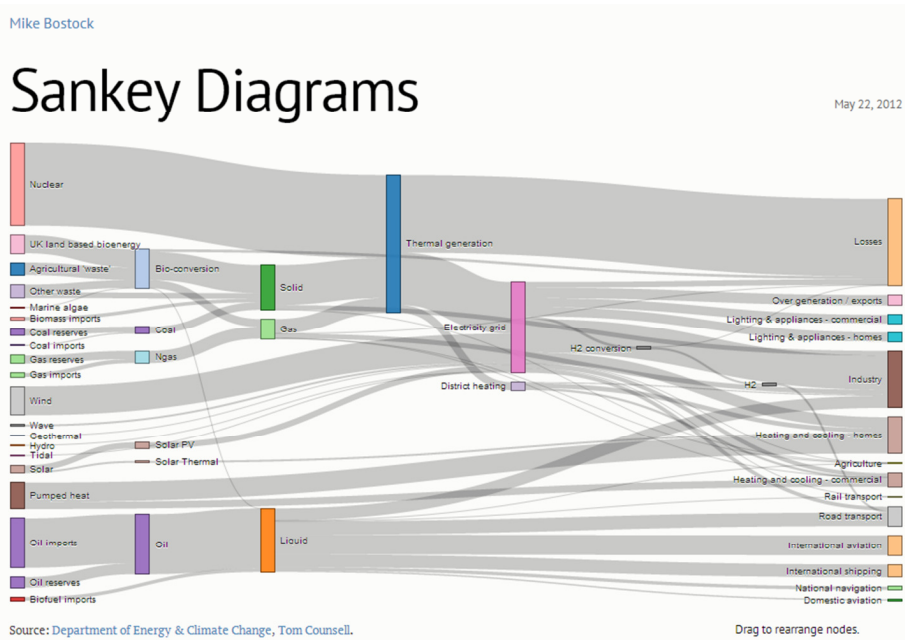

*Figure 2: Sankey Diagram originally developed for the D3.js visualization package. A Sankey diagram visualizes flow between steps in a process. In this case, the production of energy from different sources, and where it is being used (source: Bostock [5]).*

## 3. Previous work

Previous work on in-game economics is diverse, stems from both academia and the game- and associated industries, reflecting the underlying diversity in games and the economic models that are utilized across game genres.

In this section the focus will be on drawing the general outlines of research on in-game economies, but in the interest of maintaining the scope, with a focus on the previous work most relevant to the study presented here, i.e. in-game trading in MMOGs. This section is not intended to be a full review of all work on in-game economics. Overviews of different parts of this area of research- and development have previously been provided by Lehdontovirta [21], Fields and Cotton [15], Castronova et al. [55] and Knowles et al. [54], and the reader is encouraged to investigate these publications for more detailed information on topics such as economy balancing, Real-Money Trading (RMT), and the Intellectual Property (IP) regulations concerning virtual worlds. Furthermore, it should be noted that the use of simulated economies is an established research method in Economics and related fields [38,61]. Virtual worlds became interesting to economists because unlike simulations they involve real-time human behavior, within an environment that was bound by strict rules and even partially controllable.

Given the sheer scope of games that feature in-game economies, and the link to game business models, defining a succinct State-of-the-Art (SOTA) is challenging - and this is exacerbated by a couple of key complications:

1) **Recent area of work and rapid innovation:** MMOGs have been around for roughly 15 years, and it is only in recent years that the traditional retail model for driving revenue has begun to be supplemented or even replaced by other models, e.g. F2P. That being said, games have had in-game economies for decades, e.g. in the form of using earned in-game currency to buy equipment in Role-Playing Games (RPGs) such as *Eye of the Beholder* or the *Might & Magic* series. However, it was not until the advent of the MMOG that academics started working with in-game economies, with the pioneering work of e.g. Castronova [7,42]. This was mimicked by industry professionals, e.g. Simpson [32] who described the economic system of *Ultima Online*, one of the first MMOGs. MMOG-based virtual worlds, as well as non-game based virtual worlds such as *Second Life* and *There* have since then formed the basis for economic as well as social sciences research [21]. From the perspective of game development, the systems and mechanics established for controlling the flow of resources within a game environment, whether simple or complex, are foundational to the success of a game, and therefore the investigation of virtual world economics essential to the game industry [3,8,17]. With the fundamental changes occurring in the game industry in recent years, notably the drive towards social games on online platforms such as *Facebook*, and later the shift towards games on mobile platforms, and the introduction of non-retail based business models, a renewed interest for in-game economics has emerged [13,23,26,30,54,56,57], although the focus is different than the MMOG-related work typical of the early 2000s. Irrespective of these developments and the rapid innovation in the area, the actual history of work being done from either industry or academia on in-game economies is relatively young. This leads to a situation where there are few established theories or models to compare e.g. the *Glitch* data to, and lends emphasis to the necessity for more knowledge building, e.g. via explorative work focused on describing in-game economies and how players operate within them, as well as comparing these with real-world economies.

2) *Lack of data:* A concrete challenge for research into in-game economics – whether industrial or academic - is that it is highly uncommon for game companies to share financial data with research institutions or other stakeholders, due to the sensitivity of such data [11,29,54]. Research done in the industry thus stays confidential, and research in academia is hampered by limited data access. These factors combined means that the publicly available knowledge about in-game economics is limited and sometimes even anecdotal rather than empirical in nature. Furthermore, due to the proprietary nature of the behavioral data involved, knowledge exchange in the industry is also limited. This leads to a situation where the area of game-based economic analysis is fragmented and the SOTA is difficult to describe. There are indications of changes in this pattern though as there are increasing numbers of academic articles on in-game economics that are based on collaborations with game companies. For example the work of *The Virtual Worlds Observatory* (http://129.105.161.80/wp/), a collaboration between multiple universities which has notably been working with Sony Online Entertainment around the MMOG *EverQuest II* [59].

3) **Two communities, different approaches:** There are two main sources of knowledge about in-game economies: industry sources and academic research. Knowledge in both domains tends to be fragmented and there is a lack of standards and reviews, with a few exceptions [e.g. 7]. While university researchers have so far focused mainly on MMOGs, industry professionals have covered a wide swathe of games in articles, blog posts, reports, white papers and conference presentations [e.g. 58], covering not only MMOGs but more recently focused on the economics of F2P games and the social/mobile area [e.g. 15,17,35,50,51]. The approaches and goals used in these two communities also differs: Knowledge from the industry is generally focused on reports, presentations and blog posts sharing experiences and analysis from a single game or sets of recommended rules [e.g. 32,50,51]; while work from academia is typically focused on testing and evaluating economic theories within game environment [e.g. 7], describing economic systems [42] or conversely more exploratory research on e.g. human behavior. Game analytics work in the industry can integrate any source of business intelligence connected to game development, including e.g. market analysis, benchmarking, internal analysis, and customer analysis, but has notably in the game industry become applied to analyze and derive insights from player behavior towards informing and improving game design/user experience and to assist monetization [15,23,29,50,51].

The above caveats in mind, it is safe to say that in-game economies and economies around games (e.g. via RMT), vary in complexity, operation and design.

For example, a Free-to-Play (F2P) mobile game can rest its entire economy on a few items or boosters which the user can purchase from an in-app shop [15,51], with limited or no trading possible between players, leading to a relatively simple economic system of player-driven generation or purchasing of volatile or short-term assets which provide a temporary gain to a specific player (e.g. "boosters" in *Candy Crush Saga*) but does not affect other players. Similarly, a traditional Role-Playing Game (RPG) such as *Baldur's Gate* or *The Witcher* can include players generating assets as they play the game, and expending these assets on items and character improvements. RPGs can also feature object creation systems with gradual unlocking of higher-tier items. However, traditionally RPGs and other games sold under a retail business model do not include the ability for players to spend real-world money to purchase in-game assets. This is however rapidly changing, for example via Downloadable Content packs (DLC) for games such as the *Mass Effect* series.

In comparison, a major commercial, subscription-driven MMOG can operate with a complex, multi-layered economic system that provides a variety of ways for players to generate and spend resources or assets, multiple soft and hard currencies, potentially including direct purchases using real-world currency, real-world to virtual-world trading systems, and also contains a variety of sinks draining the virtual game world of resources. Combined, such mechanics serve to keep the economy in balance [17,30,54,55,56,57], even when faced with illicit player activity such as bot-driven gold farming [1,19]. To complicate matters further, the revenue models of notably mobile games but also MMOGs are intrinsically tied in with the in-game economy of these games. Thereby enters the discussion different revenue models (e.g. F2P vs. subscription-based MMOGs), and how to interlace them with virtual economy mechanics [e.g. 57].

## 3.1 MMOG economics

While game economics work in the industry is increasingly applied in the F2P/mobile contexts, academic game economics research has focused on MMOGs and other avatar-mediated massively multi-user environments, both of which are referred to as *virtual worlds* [37]. The current work is based on economic data from an MMOG, and we therefore turn our focus to this type of digital game and similar persistent, massively multi-user virtual environments.

From the perspective of academic research, in-game economies in virtual worlds are of interest for a variety of reasons, but perhaps most importantly because MMOGs and other forms of massively multi-user virtual environments form semi-controlled environments that allow investigation of human behavior – and associated economic modeling – on a societal scale. Previously, behavioral economics has been largely limited to building simulations [26]. This also in a games context due to a lack of access to controlled environments or access to large-scale datasets [11-13,19].

Formally, in the definition of Knowles et al. [54], a **virtual economy** comprises (direct quote):

1) *"The set of currencies, goods, and services – collectively, the virtual assets –that users can collect, use, and/or trade with one another in a virtual world;*
2) *The productive technologies available to users in that world; and*
3) *The markets where users execute trades with each other or the owner of the virtual economy."*

With the addition to the first point of virtual assets that users can buy or otherwise produce – either for in-game currency (soft or hard) or for real money, this definition appears to be valid for both simple and complex in-game economies as well as RMT. Not all virtual economies will feature all of these features, but they appear to be common in MMOGs, including *Glitch*, and the definition of Knowles et al. (2013) is therefore adopted here.

In *Glitch* as well as many other games where the operator (game company) of the virtual environment in question permit the players to engage in some sort of **trading** – irrespective of the complexity of the trading system -such trade is governed by rules. These rules can encompass for example trading volume, define what items, currencies or services that can be traded and how, and the types of exchanges that are permitted. For example, in *World of Warcraft,* not all items can be traded, and there is no operator-sanctioned system for trading in-game virtual items with real-world money, although third-party operators facilitating this exist. In *Glitch* users were allowed to trade most items generated in-game, with the operator taking a cut of the proceeds (Figure 3, and see Section 4, below).

It is common in MMOGs that venues facilitating trade between players or between player and the operator are provided, e.g. an auction house or a bazaar. The nature and rules surrounding these varies, and sometimes economic forces drive players to designate spaces for trading as noted by Castronova [42]. In-game or across-game trading systems such as auction houses, do not contain any information asymmetries between buyers and sellers like real-world auction sites such as *eBay*. Though there are vast differences between different kinds of items in *Glitch,* all items of the same kind are completely identical. Furthermore, information about items is freely available (viewable by "mousing over" the item or selecting it). Added to this the ability of a MMOG developer to also track and analyze behavioral data and demographic data about the player population, has opened up new ways for economists to study consumer behavior [e.g. 8,13,55].

**3.2 MMOG trading – key related work**
Having outlined the general features of the work on virtual economies from academia and industry, and the basic features of MMOG economies, we turn to describing some of the key work related to the trade analysis and –visualization presented here.

In-game economics have been approached from a variety of angles, including testing real-world economic theories in virtual environments [e.g. 8,43,44], interaction between real world and virtual world economies [37,64], notably in terms of assets trade, or focusing on real world economies themselves [e.g.

60]. Legal theory has also been applied to investigate the trade of virtual assets and debate ownership of virtual assets.

Several authors have examined auction house data from MMOGs specifically, including Simpson [32], whose early work on *Ultima Online* (one of the first MMOGs) described an environment with hundreds of different items being exchanged via virtual auction houses, and highlighted the macroeconomic design built into the game from the onset to facilitate economic flow. Importantly, Simpson [32] highlighted the failure of a shopkeeper-run (not player controlled) economy, which led to *Ultima Online* experimenting with player-run auctions, i.e. more open economic systems, forming the foundation for this feature in future games.

More recently, Castronova et al. [8] investigated macroeconomic behavior in *EverQuest II* working with transaction data, proposing an empirical test of whether aggregate economic behavior maps from the real to the virtual world. The authors noted that an aspect of virtual world auction data that makes adoption of real-world aggregates in the virtual domain easier, is that items listed for sale in *EverQuest II* are simple to map to real-world analogs, e.g. food items, furniture items, etc. This is also the case for e.g. *World of Warcraft* and *Glitch*. Morrison and Fontenla [24] studied eight *World of Warcraft* servers, and noted that price convergence occurred at all eight of these. Finally, Lohdonvirta [21] used virtual item sales records to identify attributes that drive purchase decisions.

From the perspective of economics research in academia, most current work has focused on mapping existing concepts onto the games – as noted by Y. Varouakis in Plumer [26]: *"They're just imposing their own prejudices rather than genuinely trying to extract new knowledge from these worlds"*. Given the unique nature of MMOGs, integrating game design in economic studies was advocated by Castronova et al. [8], who suggested that "code is law", i.e. that macroeconomic outcome in a virtual world is largely explained by design structure, the parameters of which can be different than the real world, and thus care must be taken when integrating real-life economics theories and -models into the domain of game worlds. Williams et al. [46] reported differences in the representations of e.g. gender, race and age in games as compared to real-world populations. This is reflected by Szell and Thurner [65] who noted that although recent research has provided evidence that human behavior inside game societies might mimic real-world communities [e.g. 45], this remains an unresolved issue.

It is important to note that previous work on analyzing virtual economies has often been hampered by a lack of direct data access, leading to most work being done using secondary data, e.g. scraped from websites, surveying players or via mining the client-server data stream. In essence, due to the sensitive nature of financial data from in-game economies, the current knowledge base is highly fragmented and there is a lack of data-driven knowledge [11,26]. Exceptions include Castronova et al. [8], who obtained access to data from the MMORPG *EverQuest II*, and used this to investigate and calculate metrics for e.g. production, consumption and money supply based on real-world definitions. Additionally, Szell and Thurner [65] compiled data from 300,000 players of the MMOG *Pardus,* focusing on social network analysis across multiple types of networks in an environment where players engage in virtual economic activities of various kinds.

### 3.3 Summary
MMOGs provide virtual environments that facilitate socialization and interaction across a range of vectors, and the motivation for players to participate in MMOG play is equally diverse, covering e.g. competitive play, establishing social connections, gaining respect and status within a virtual society, exploration as well as fun and entertainment, and even to destroy the work of others [16,40-42,44].
MMOG economies are complex systems that have to be carefully designed and managed to ensure that they properly support the population and provide useful functionality for generation, exchange and elimination of resources [3,4,8]. This is indeed the case for any online game with an integrated economy [15,29]. In conjunction with the social and combat mechanics of an MMOG, these systems form the backbone of these games. From the perspective of game development, MMOG economics are vital and has in recent years seen professional economists getting employed with MMOG (and other game-) companies to design and manage the complex and dynamic economies of virtual worlds [see e.g. 17,26,30].

## 4. Glitch

*Glitch* was a browser based Massive Multiplayer Online (MMO) by San Francisco based startup Tiny Speck that ran from 2011 to 2012. The majority of the games lifetime was in beta, with a short stint as public release from September 27, 2011 to November 30, 2011. *Tiny Speck* announced on November 14, 2012 that the game would close December 9, 2012. At the time of public launch the game had 27,000 active users and a free to play model with no transactional elements in place [33].

*Glitch*'s core gameplay revolved around crafting, resource gathering, trading and social elements in an open-ended world. The main objective of the game was to help build the world and create mini-games within. From the art style to the gameplay the driving idea behind glitch was an imaginative player driven ever expanding universe, comparable to a more stylized version of *Second Life* (Figure 1). Structural elements like leveling and questing existed to facilitate the acquisition of resources needed to fully participate in the world-building mechanics.

The in-game currency, Currents, were obtained through questing, various locations throughout the world, selling items to NPCs, and selling items to other players in the in-game auction 20]. Similar to other in-game MMO auctions, players could post any quantity of an item on the market and see other player's posting of the same items: *"Postings expired after three days, and Tiny Speck would claim a small fee for each posting. Initially listing an auction cost a 3% Currant fee, followed by an 8% commission fee (in Currants). On May 25, 2012 commissions on auctions were dropped and listing fees were adjusted upward to 7%"* [20]. Third party developers also created tools for users to track items in the auctions house when away from the game via smartphones and tablets. The auction house was not the only means of in game transactions. Players were also able to privately trade with one another and bypass the auction completely.

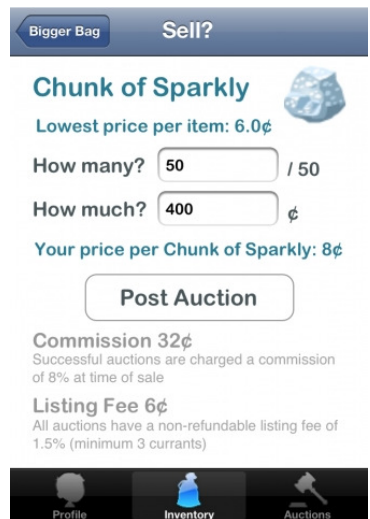

*Figure 3: A screenshot showing a player posting an auction in Glitch.*

## 5. Data

During the lifetime of the game, from shortly following launch until the game shut down, Landwehr [20] collected telemetry data from the game. The data collected focus on auction house behavior, notably sales between players and records of item prices in the game, but also other data, e.g. on friendships, which is however not used here due to lack of temporal information. Landwehr [20], after collecting the data, wrote a brief technical report describing the data, and afterwards released the data in the public domain in May 2013 [for download instructions see Landwehr [20]).

## 5.1 Data collection

During the relatively brief life of *Glitch*, thousands of players engaged in financial transactions via the games native auction house (Figure 3). The data contain an account of the behavior of the players in relation to trade that cross the lifetime of a MMOG. Such a dataset have not previously been available for study outside of the companies that run and manage MMOGs. The collected data centered on in-game auction and economic activity, as well as general forum discussions and friend networks. Using the game's built-in API for $3^{rd}$ party developers, Landwehr [20] scraped the data using a mix of Python scripts and Curl HTML scraping tools to mine four key areas of data: auction sales data, item street prices, forum conversations, and in-game friendship networks. Data is publically available in CSV, HTML XML formats. Due to do the lack of temporality in the friendship data and street prices data, we excluded this data from our analysis.

The auction sales data was retrieved via a Python script that pulled the list of active auctions one every 10 seconds into a MySQL database from November 17, 2011 to December 10, 2012. Landwehr [20] noted that several intervals were lost due to disk space/memory issues. Thus the data set cannot be considered completely comprehensive of every in-game auction posting. Approximately 3 million auctions were collected over 14 months. For every auction posted the following data fields were collected: player id, timestamp, action expiration date, item name, item category, item quantity, tool uses, tool capacity, and the final outcome of the auction (sold, expired, or deleted) (see Landwehr [20] for more detailed data dictionary)

The data scraped from the forum was provided in two forms. A CSV table of all forum posting was provided in two forms: 1) a CSV file of player id, timestamp, and comment index number 2) separate HTML files corresponding with each comment index with the forum posting text. It is important to note that some posting were lost due to the nature of actively scraping the forums while in use [20].

## 5.2 Daily Activity

At the peak of activity, *Glitch* had 8357 Daily Active Users (DAU) viewed on a monthly basis, gradually reduced to 63 at last month the game was active (Figure 4). Two spikes in activity occurred during September and October, although - to the extent it has been possible to investigate the causes of this increase in the active users - this appears to be unrelated to e.g. content updates or media coverage of the game.

| | | | | | | | MONTH JOINED GAME | | | | | | | | | total players | % of total |
|---|---|---|---|---|---|---|---|---|---|---|---|---|---|---|---|---|---|
| | | Nov-11 | Dec-11 | Jan-12 | Feb-12 | Mar-12 | Apr-12 | May-12 | Jun-12 | Jul-12 | Aug-12 | Sep-12 | Oct-12 | Nov-12 | Dec-12 | | |
| Number of Months Played | 1 | 1220 | 1432 | 452 | 243 | 166 | 231 | 227 | 164 | 102 | 871 | 916 | 1038 | 1198 | 97 | 8357 | 41% |
| | 2 | 1454 | 690 | 222 | 107 | 104 | 112 | 103 | 60 | 55 | 420 | 529 | 741 | 153 | | 4750 | 23% |
| | 3 | 898 | 378 | 131 | 78 | 62 | 58 | 60 | 28 | 25 | 223 | 325 | 148 | | | 2414 | 12% |
| | 4 | 597 | 268 | 88 | 46 | 38 | 32 | 19 | 24 | 8 | 166 | 64 | | | | 1350 | 7% |
| | 5 | 462 | 188 | 64 | 43 | 36 | 31 | 17 | 8 | 7 | 34 | | | | | 890 | 4% |
| | 6 | 376 | 148 | 47 | 22 | 18 | 7 | 11 | 5 | 4 | | | | | | 638 | 3% |
| | 7 | 282 | 131 | 23 | 10 | 13 | 8 | 9 | 2 | | | | | | | 478 | 2% |
| | 8 | 236 | 84 | 21 | 6 | 1 | 9 | 1 | | | | | | | | 358 | 2% |
| | 9 | 213 | 63 | 19 | 5 | 8 | 2 | | | | | | | | | 310 | 2% |
| | 10 | 204 | 46 | 18 | 3 | 3 | | | | | | | | | | 274 | 1% |
| | 11 | 129 | 24 | 8 | 3 | | | | | | | | | | | 164 | 1% |
| | 12 | 94 | 19 | 2 | | | | | | | | | | | | 115 | 1% |
| | 13 | 90 | 15 | | | | | | | | | | | | | 105 | 1% |
| | 14 | 63 | | | | | | | | | | | | | | 63 | 0% |

*Figure 4: Distribution of Daily Activity*

## 5.3 Auction house activity

For our analysis, we focused on the in-game auction activity using auction sales data and forum posting related to the game market place, as a proxy to better understand overall game health in relations to player activity. There were 2,914,359 data points on the auctions table, where we see 20,266 unique players listing 679 unique products across 41 unique categories from 11/17/2011 to 12/10/2012. 80% of total auctions are in 10 (41 total) categories, with 62% of total auctions coming from the top 5 categories.

More granular, 28% of total auctions are in 10 products, with "meat" being the most popular auction item - 8% of all auctions.

There was an average (μ) = 7,472 auctions per day with a standard deviation (σ) = 4,081. The high was seen on 12/04/2011 with 25,252 auctions created, while the low was during the final days of data collection on 12/10/2012 with 318 logged. At a monthly level, there was an average of 208,168 auctions with a standard deviation of 108,928. December 2011 marked the high with 505,873 auctions listed and December 2012 being the low with 29, 478 logged through the first 10 days of the month (Figure 5).

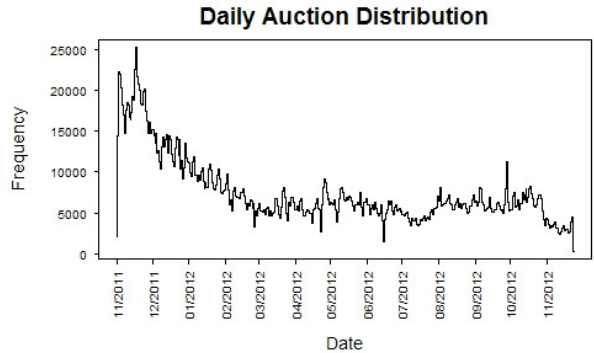

*Figure 5: Distribution of daily auction activity*

On average, there were 143 auctions listed per player participating in the auction house, with σ= 546. The range among the players was from 1 to 2,133. Finally, we saw μ = 85% of auctions successfully resulting in a sale (σ = 2.9%) and μ = 35% success per player per month who posted (σ= 47.7%) (Figure 6).

The data scraped from the forum was provided in two forms: 1) a CSV file of player id, timestamp, and comment index number 2) separate HTML files corresponding with each comment index with the forum posting text.

For our analysis, we focused on the in-game auction activity using auction sales data and forum posting related to the game market place, as a proxy to better understand overall game health in relations to player activity. There were 2,914,359 data points on the auctions table, where we see 20,266 unique players listing 679 unique products across 41 unique categories from 11/17/2011 to 12/10/2012. 80% of total auctions are in 10 (41 total) categories, with 62% of total auctions coming from the top 5 categories. More granular, 28% of total auctions are in 10 products, with "meat" being the most popular auction item - 8% of all auctions.

There was an average (μ) = 7,472 auctions per day with a standard deviation (σ) = 4,081. The high was seen on 12/04/2011 with 25,252 auctions created, while the low was during the final days of data collection on 12/10/2012 with 318 logged (Figure 6). On average, there were 143 auctions listed per player participating in the auction house, with σ = 546. The range among the players was from 1 to 2,133. Finally, we saw 85% of auctions successfully result in a sale, with μ = 85% of successful auctions per month (SD 2.9% points) and μ = 35% success per player per month who posted (σ= 47.7%) (Figure 6).

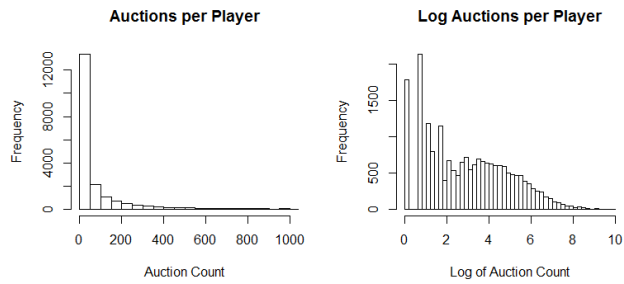

*Figure 6: Frequency distributions of auctions per player. Left: Raw frequencies. Right: Log-scale frequencies.*

### 5.4 Item valuation

As part of the data set made publically available by Landwehr, street prices of items sold were scraped from the game. These street prices represent what a vendor (NPC) might charge for a good. Instead of going to the auction house to sell an item, a player could sell an item to the vendor and receive 70% of the street price in currants (the in-game currency). *Tiny Speck* would periodically update these street prices. At the time the data was scraped on December 11, 2012, street prices were available for 560 of 742 items. Items not listed in the street prices were either items that had been removed from the game or items or rare items that were never classified by *Tiny Speck* [20]. We compared the last 4 months of cost per unit sales prices (September 9, 2012 to December 9, 2012), with the final scraped price on December 11, 2012. We are assuming that *Tiny Speck*'s street price valuations did not deviate during that time period.

Overall, item sale prices exhibited skewness, largely due to high sales outliers. Of the 506 unique items sold in the last 4 months of the game, 30.6% exhibit strongly right skewed distributions. For example, hooch (classified as drink), the 3$^{rd}$ most popular item by auction posting frequency, exhibited periodic spikes in which the item would sell for 1 million currants, which otherwise sold for 9. (Figure 7)

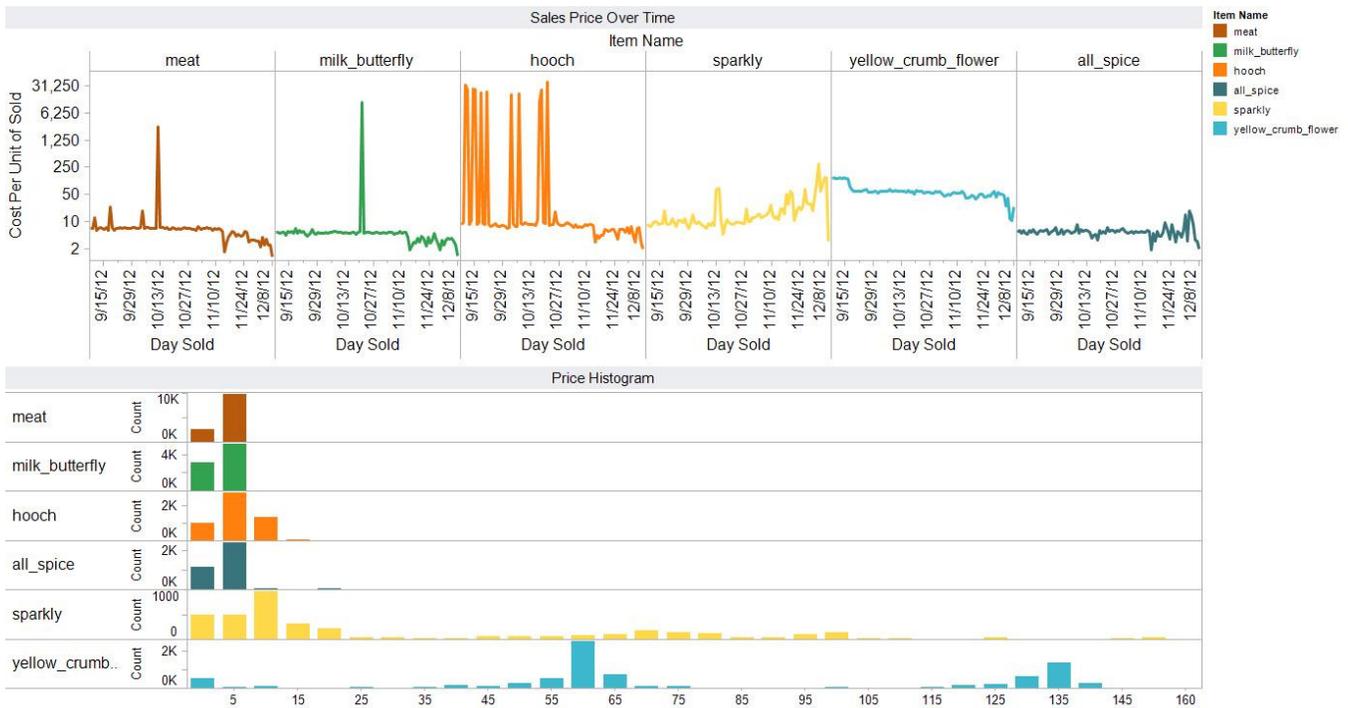

*Figure 7: Temporal distribution and histogram of final selling price using top 6 most frequently posted items as an example*

From a temporal distribution perspective, 41% of items depreciated in valued (especially in the last 4 weeks of the game) as in the case with meat, milk butterfly, and yellow crumb flower. 4% of items appreciated in value over time. In the last four weeks of the game 50% of the items depreciated in overall value (as measured by sales prices) and 12% increased. There was more more market volatility in the last 4 weeks of the games, mostly likely as a result of the announcement on November 14, 2012 that the game would be ending in December 2012.

Looking at the final street prices, we wanted to know how frequently an item sold for less 70% of the street price value (the amount a vendor would pay for the item). If a players sells an item in the auction for less that what they could sell to a vendor, the auction house may lose value for players as a viable place to sell their goods rendering this part of the game largely ineffective for player advancement in wealth. From the last 4 months of *Glitch*'s auction data, 59.1% of items sold above their street price more than 50% of the time, presenting the auction house as a more lucrative sales channel over vendors. (Fig. 8.1). We looked at the frequency an item sold at greater than 70% of its street price value. We called the proportion of this figure to all sales, the success. Figure 8.1 shows the success ratio for all items sold in the last 4 months which were also available to sell through a vendor. It is interesting to note that many of the most frequently sold items such as *meat*, *butterfly milk*, and *hooch* has success ratios below 50%. For example, *meat* sold above the vendor price 25%. Players may generate higher return on their investment if they avoided the auction house with this item. (Fig. 8.2). When players have multiple outlets for selling items in MMOs, it is important that the developers monitor and regulate these sales channels closely to maintain the economic balance and avoid negatively impacting the gameplay experience [54].

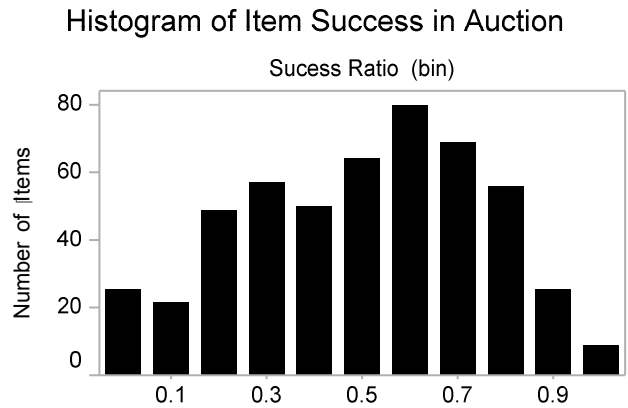

*Figure 8.1: Histogram of Item Auction Success as measured by proportion of auctions sold above the vendor price.*

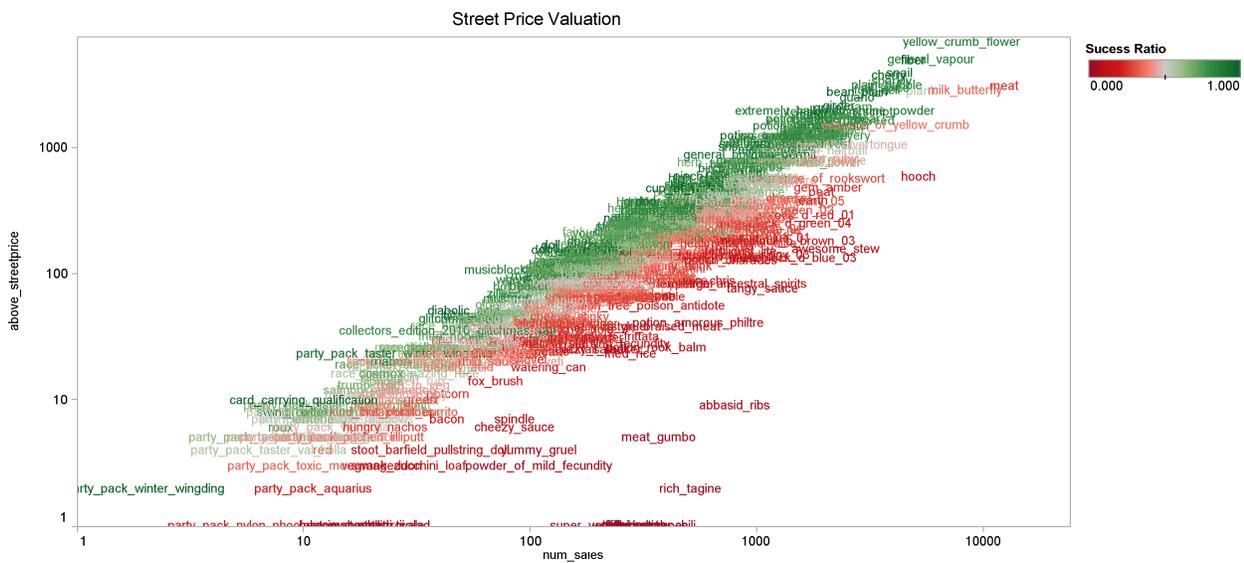

*Figure 8.2: Auction Success Ratio Valuation for all Items sold*

## 6. Methodology

The work presented comprised two components: cluster analysis and generation of the Sankey diagram. These are described separately.

### 6.1 Cluster analysis

To perform the cluster analysis 5 KPIs (key performance indicators) were defined that would be actionable for measuring the health of the in game economy: Total Auctions, Average Auctions Posted per Day, Percentage of Auctions that Resulted in Sales, Number of Distinct Items Categories Posted, Marketplace Forum Posting (Boolean). These calculated variables were aggregated from the both the auctions data table and the forum posting data table. We chose each month as a natural point for partitioning the data. Across the 14 months of data collection, players participated in the auction house μ = 2.76 months , with σ = 2.5 months). 41% of the players participated in the auction system for only one month (Table 1).

After testing for independence, we applied a clustering algorithm to each of the 14 months' worth of data. To compensate for right skewness in the data, input variables were standardized into quintiles. To determine the clusters for each monthly interval we used the Hartigan and Wong centroid seeking algorithm (K-means) [52]; for accessible explanation see Hastie et al. [53]. K-means attempts to create non-overlapping spheres around each cluster center with a greedy heuristic approach that maximizes heterogeneity across clusters and minimizes homogeneity within clusters. Similar to other greedy heuristics, K-means selects an initial set of cluster centers defined as starting seeds. $\hat{\mu}_k^0 (p \times 1)$, where k = 2…12 representing the number of cluster centers. Given the size of the data and the relative low complexity of computing the algorithm in R, 100 random starting values were selected from a specific starting seed (to more easily replication results across runs). For each iteration, the Euclidean distances is measured from each observation ($i$). $y_i^h = argmin_k \| x_i - \hat{\mu}_k^{h-1} \|^2$, where $h$ = the iteration counter. The cluster centers are then computed as $\hat{\mu}_j^h = mean(x_i : y_i^h = k)$, where j represents clustering variables, followed by the SSE for each cluster generation:

$$SSE = \sum_{i=1}^{n} \left\| x_i - \hat{\mu}_{y_i^h}^h \right\|^2.$$

This process is repeated until the squared sum of errors (SSE) is minimized and cluster assignments are consistent. Using K-means does not guarantee convergence at a global minimum, largely due to its sensitivity to starting seed values. This happened in later months with smaller data sets when trying to at a solution for the more extreme bounds of clusters (i.e. 2 or 12).

We determined the optimal set of cluster through scree plots of the SSE within each of the clusters (W) and between the clusters (B) with the intent to minimize the ratio of W/B such that the marginal benefit of (W/B) for each subsequent cluster size is minimal, where $W = \sum_{i=1}^{n}(x_i - \hat{\mu}_{yi})(x_i - \hat{\mu}_{yi})'$ and $B = \sum_{i=1}^{n}(\hat{\mu}_{yi} - \hat{\mu})(\hat{\mu}_{yi} - \hat{\mu})'$. Thus we attempt to find the most distinctive homogenous clusters. For consistency across months, since each month was cluster independently while arriving at similarly formed clusters, we sought a W/B ratio of 0.3

While the number of active in-game auction house users diminishes over time, the relative size of each of the clusters, describing specific patterns of user behavior, remains relatively constant (Table 1). For example, over the 14 months of hardcore players represent 22% to 28% of the total population at any given month. Among those clusters that are manifest for the majority of the months, the Moderate Farmer cohort exhibits the greatest variance in cluster size across the months, while Forum posts are most consistent in their population size from month to month.

| Cluster | μ cluster size across months | σ | Months manifested |
|---|---|---|---|
| Moderate | 0.16 | 0.00 | 2 |
| Forum | 0.04 | 0.01 | 13 |
| Hardcore | 0.24 | 0.02 | 14 |
| Casual Losers | 0.14 | 0.02 | 12 |
| Casual Winners | 0.22 | 0.02 | 12 |
| Moderate Farmers | 0.15 | 0.03 | 14 |
| Moderate Losers | 0.14 | 0.02 | 6 |
| Moderate Miscellanea | 0.16 | 0.02 | 9 |
| Causal Forum | 0.03 | 0.00 | 4 |
| Moderate Winners | 0.15 | 0.00 | 1 |
| Casual | 0.35 | 0.00 | 2 |

*Table 1: Cluster sizes and number of months manifested.*

## 6.2 Player profile descriptions

In full, four main player types emerged, Casual Players, Moderate Players, Hardcore Players, and Forum Players (Fig. 9). Within these larger types clear distinctions emerged, resulting in 11 distinct player types, discussed below. Those that remained at the most basic type (i.e. Casual, Moderate, Hardcore) occurred in months where there were too few players to create divergent clusters each month (these two months contained only 5 groups).

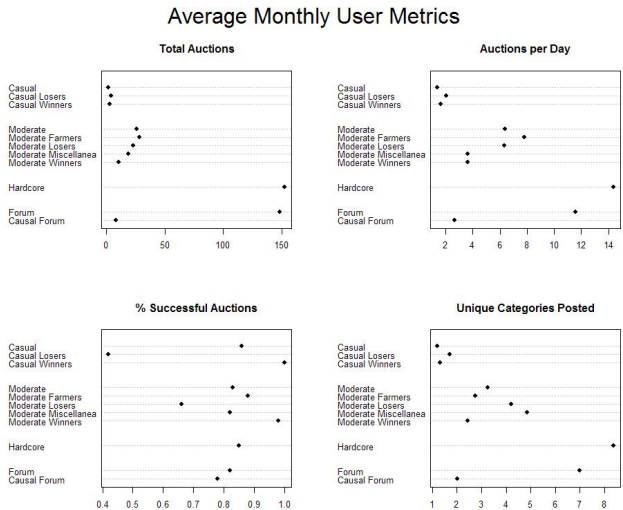

*Figure 9: Average monthly values for each of the input variables in the cluster analysis grouped by cluster. While standardized values were used for the clustering the above chart represents the raw average for direct comparison.*

**Casual Players:** Among casual players, two groups exist with key distinctions by auction success rate. Casual Losers have 42% sales rate on all auctions, while Casual Winners have 100% success rate (i.e. all auctions posted are sold).

**Moderate Players:** Moderate players are characterized by mid-range activity level. Moderate Farmers post a large number of auctions per day across a smaller range of item categories. These players are focused and establishing a trade niche within the economy. Moderate Miscellanea are less focused, instead posting auctions across a broader range of categories at a more leisurely pace (i.e. fewer auctions per day). Moderate Losers post more frequently than Moderate Miscellanea but are less successful than all other clusters in the moderate category.

**Forum Players:** Forum players have a similar economic play style to that of hardcore players with the exception of their posting activity in the marketplace forum. During the months of Feb – May 2012, a group of more casual players begin posting in the forums.

**Hardcore Players:** Hardcore players exhibit the least divergence in their group and across time. These players perform at the upper bound of all KPIs.

## 6.3 Sankey diagram design and development

Sankey diagrams comprise a series of nodes and links, similar to network diagrams (Fig. 10). The key difference is that nodes are positioned or stacked at key intervals along one of the axes. For instance, at month t, players are categorized into 5-7 player behavior clusters. These players may have shifted clusters (or entered or left the game) from the previous period and into the next period. Nodes are positioned at months *1,2,…,14* while links show the shifts in players across clusters from one month to the next.

The diagram (Figure 9) was developed using D3.js (Data-Driven Documents), a javascript library for binding data to HTML elements and Scalable Vector Graphics (SVG) shapes to dynamically render data visualizations (Bostock [5], see also Tamc [63]). For an introduction on how to use D3.js in mobile analytics, see Seufert [50]. Scalable Vector Graphics (SVG) are a set of browser rendered shapes that scale appropriately to different resolutions and can offer interactivity. The scalability of SVG objects differentiates them from other pixel-based image formats.

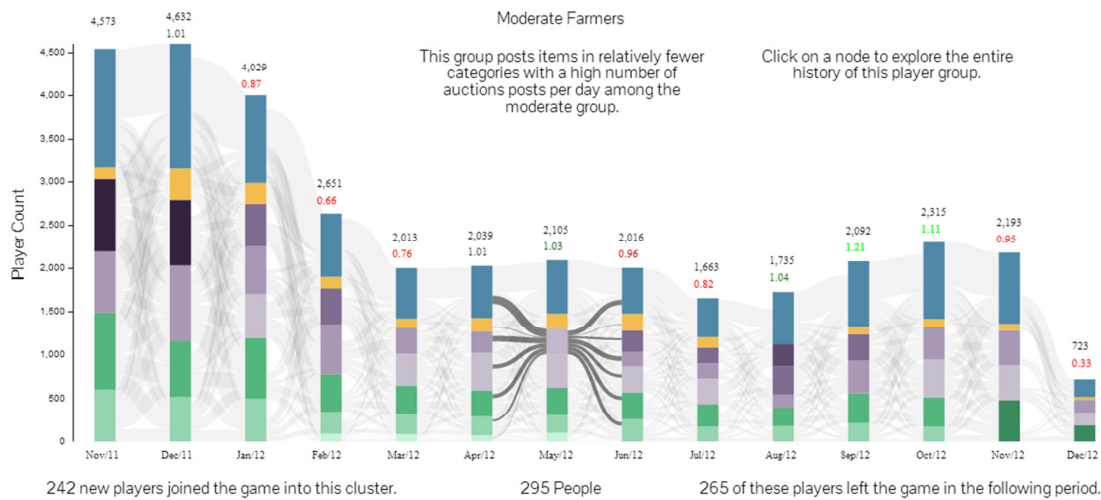

*Figure 10: Glitch Sankey Diagram: Combining the click through elements of the Sankey diagram with standard bar chart formatting and measurement*

Utilizing the Sankey diagram plugin for D3.js (http://bost.ocks.org/mike/Sankey/), we reduced any padding, the amount of whitespace, between the vertically-placed nodes to replicate a stacked bar chart at any time *t*.

Hovering the mouse over any link or node shows the respective player count, cluster descriptions, and information about new and departing players. Total player count, a one-month retention rate, and x-axis labels were added via an in-file javascript variable. Additionally, a click-through of any node will provide a more detailed view of that group of players through the 14-month history (Fig. 11).

The raw data to be plotted was a javascript object notation (JSON) file containing separate node and link information. A JSON file is a hierarchically structured document describing objects through the use of parents (objects) and children (descriptors). A given parent can contain multiple children and sub-children, and various data types can be exist in a single file. Our data contained two parents: nodes and links. The nodes parent contained children of month, cluster mapping, value, color, and departing and joining player counts, while the links children were source, target, and value. Links were mapped from source to target (with curvature defined in the Sankey.js file), and the stroke width was sized by the value of players moving from source to target. The nodes were sized based on the value noted in the JSON file, representing the size of players existing in that cluster in that month. We also adjusted the order of the raw data file such that we maintained the same order of clusters within each month, helping to better organize the visual and reduce overlap of the links from node to node.

Without adjusting the Sankey.js file, which contains all the code for computing the node depths, link connections, etc., there are a few easy to adjust arguments when calling the Sankey() function: node width, node padding, and layout. We've noted previously that padding refers to the space between nodes, width refers to the width along the x axis of the rectangular nodes, and layout refers to the y-axis positioning. To reduce the Sankey diagram down to the stacked bar graph, we reduced the padding to 0 and also changed the layout to 0. The 0 layout will draw the objects from the top of the SVG canvas and build downwards. In order to draw an SVG shape, first a canvas must be specified or created if not existing. Shapes are then drawn on top of the canvas just like the process of any brush and canvas physical painting. In order to base the overall graphic along the bottom of the SVG canvas, we utilize the SVG transform attribute, which controls positioning, to flip the SVG canvas through the use of two specific inputs: translate (x and y movement) and scale (inverting/flipping image). With this positioning adjustment, the position of each of the major elements entering the screen also needs to be transformed.

These positioning changes to the default library added a few advantages to the public examples. By reducing padding and basing all bars to the bottom of the canvas, we can better compare respective cluster sizes over time as opposed to the default layouts which may require repositioning of nodes though click and drag interactions.

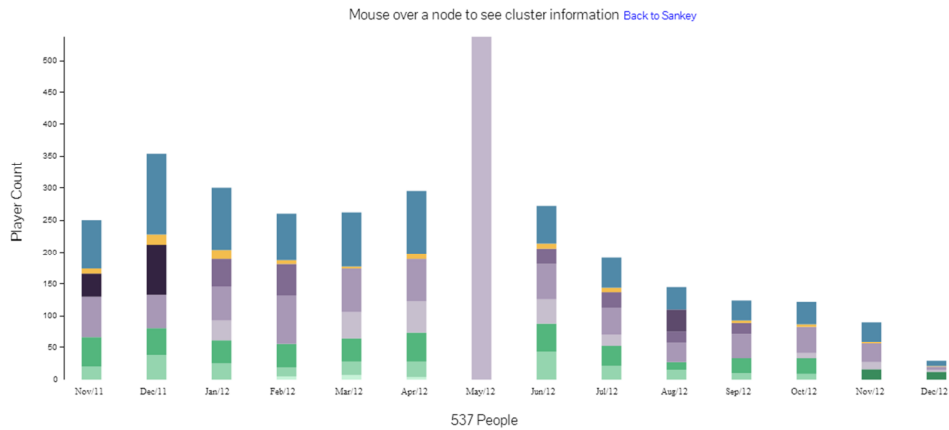

*Figure 11: Visualization on cluster click through: clicking on a node reveals the distribution that lead up to the node and the distribution in proceeding months for said node.*

**7. Key results**
A variety of results are evident from the analysis and visualization presented. Key results will be presented in this section, and discussed in further detail in the next section, as follows:

**Glitch economy:** The results show that the auction house of *Glitch* was used by the majority of the players, with relatively great success (85% average success chance of making the sale), although the majority of the players infrequently used the auction house. Furthermore, the use of the auction house followed a limited set of patterns, as exemplified by the clusters. The number of daily auctions dropped rapidly during the initial months of the game's lifetime, but more or less stabilized thereafter, despite background changes in the number of players, before finally dropping rapidly in the last month the game was active. Finally, the majority of the auctions were for a small subset of the items represented in auctions, with 62% coming from the 5 most frequent categories.

**Player behavior over time:** The analysis clearly indicates variations in the player behaviors over the duration of the game, despite the consistent presence of the four high-level clusters throughout the lifetime of *Glitch*. Taking into account the strong overall economic fluctuations over time (diminishing player counts and auctions per player); someone who was a moderate player in December 2011 is almost equally likely to fall into the any of the core clusters. Furthermore, players do not remain in single cluster for the duration of their lifetime. Players move relatively freely across clusters from month to month. Overall, players do not remain in the same cohort throughout their game life. Players who participated in the auction house for 7 or more months (half the total available months), moved between 4.4 clusters in their lifetime (Table 2). A similar pattern of changes in playstyle was also documented by Sifa et al. [31]

in the game *Tomb Raider: Underworld* – an entirely different game than *Glitch* (a first-person action-adventure game vs. an MMORPG). The authors used Simplex Volume Maximization to define clusters of behavior, based on 11 performance indicators tied in with the core game mechanics (jumping, combating enemies, etc.), as a function of the game's levels. The results showed that the relative proportion of people adhering to specific playstyles varied throughout the game, and that new playstyles emerged and disappeared on a running basis.

| Number of Months Played | Average Number of Distinct Clusters |
|---|---|
| 2 | 1.72 |
| 3 | 2.32 |
| 4 | 2.80 |
| 5 | 3.29 |
| 6 | 3.63 |
| 7 | 3.94 |
| 8 | 4.25 |
| 9 | 4.53 |
| 10 | 4.82 |
| 11 | 4.89 |

*Table 2: Average number of distinct clusters as a function of numbers of months played.*

**Casual and hardcore players:** Taking an in-depth look at the individual cluster nodes, players are most likely to enter and leave the game as a casual player, resulting in the highest inflow/outflow as a percentage of the cluster compared to other clusters. We can speculate that this is where players first get acclimated to the auction system as a casual player, and as their interest in the game wanes they returned to a casual status. Additionally, looking at the full distribution of casual players at any given month showed a dearth of players near end-game months (November - December 2012). For example, within the first three months there were very few players that ultimately make it to the end of the game compared to hardcore players. Hardcore players, on the other hand showed the highest proportion of consistency across clusters.

At any given month, around half of hardcore players (average) remained Hardcore players in the preceding and proceeding months. The insularity of the segment was higher than that of any other group. This pattern is also indicated by reports from the game industry, that also indicate that elite players, and/or alternatively those that spend lot of money on virtual items (the proverbially mentioned "whales" segment) [15], are the least likely to leave the game (e.g. Nozhnin [39]). It should be noted however, that industry reports generally do not reveal their specific methods nor the datasets used, so results cannot be independently verified.

Additionally, hardcore players were least likely to leave the game. Especially in the early months of data tracking, Hardcore auction users and forum posters were most likely to stick with the game until the end. Hardcore players stayed in the game $\mu = 5.2$ months, Forum posters $\mu = 7.4$ months, while Casual users $\mu = 3.2$ months. This highlights the importance of users' engagement. Those who were active in the online community outside of the game had the highest retention rates. Having the tools to track these people is critical, considering moving from a Hardcore to a Casual status can thus be an indication of potential player departure (churn).

**Sankey diagram:** the Sankey diagram allows the observation of variations in player behavior over time. This can serve as benchmark for tracking disaffected players, compromised accounts, which types of players are at the highest risk for leaving the game, etc. While the process for the research presented here was manual, the methodology can be scaled up to be fully automated (see discussion Section 7, below).

# 7. Discussion: Using Sankey diagrams for progression analysis

The analysis presented here focuses on the auction house component of the economic system of *Glitch*, and combines with temporal perspectives and the use of Sankey diagrams to track and visualize the flow of players between behavioral clusters. Above results have been presented specifically concerning the in-game economy of *Glitch* during the lifetime of the game.

It is clear from the raw numbers that *Glitch* experienced a declining user base over time, but by integrating analysis with visualization, it is possible to observe the user base and auction house behaviors change over time, providing a means for tracking the temporal elements of the clusters and associated behaviors.

The methodological approach presented here is based on individual elements that are known in game analytics research or other contexts: k-means clustering of behavioral data from games and normalization of the same [12], temporal bins [31] and Sankey diagrams [27,28]. Combined, these provide a framework for analyzing and visualizing player behavior in a game that is applicable outside the specific example presented here.

When building games, a specific operational space is designed for the players. Identifying how players operate within that space is important to evaluate the effectiveness of a design and investigating if design intent has been realized. Because play happens as a function of some other dimension – time, levels, character development, narrative progress etc. – being able to map playstyles as a function of these dimensions is useful for e.g. evaluating design, player engagement and monetization strategies [13,15]. These types of analysis are thus important across both of the two main perspectives that can be applied on people who play games: as players and as customers [14].

Fundamentally, what the work presented here deals with is the combination of progression- and playstyle analysis [11,12]. **Playstyle analysis** seeks to identify patterns in how people play a game. Playstyles can be sought identified across the entire game, or just a section of it. **Progression analysis** focuses on evaluating how people progress through a game.

Both finds use in game development [29], and are also important in game AI and procedural content generation for games [47-49]. In the current case, the playstyles are defined via the behavioral clusters, and progression defined in terms of time bins (or buckets), each of a duration of one month.

Progression analysis is comprised of four fundamental components, of which the first two comprise the steps involved in performing a playstyle analysis, i.e. defining and grouping playstyles for one or more players (note: there are many ways to develop playstyles, this is just an example):

1. **Feature selection:** Choosing the right behavioral features (manually or automated).

2. **Dimensionality reduction:** Reducing the number of input features (optimization) is often necessary due to the high number of potentially interesting behavioral features in digital games. A large number of techniques are available for dimensionality reduction (e.g. clustering, segmentation, classification), all with specific strengths and weaknesses. Data normalization is a key step in most machine learning-driven processes given the varied nature of most input features [12,18].

3. **Progression mapping:** Mapping reduced/compound features to one or more progression dimensions. The choice of dimension depends on the goal of the analysis. Progression in games can be temporal (playtime), content-based, relate to narrative progression, skill progression, spatial movement, or any of these combined. Progression can also be measured in different ways, for example character wealth, level, kills, mushrooms collected etc. Multiple measures can be combined to develop compound constructs. The main challenge is selecting the right dimension or dimensions of progression.

4. **Visualization (and implementation):** Developing stakeholder-flexible interactive visualizations of results, and secondly making decisions based on them and implementing these.

In practice, playstyle and progression analysis can be combined as described above, to explore how playstyles vary as a function of a specific dimension, e.g. time. These kinds of analysis can be performed using a great variety of techniques, from basic aggregation of key behavioral features; to complex machine learning approaches [see e.g. 12,13,23,25,39,43].

The Sankey diagram model presented here provides a framework for combining playstyle and progression analysis. For example, evaluating playstyles across multiple game levels, with the goal of examining which strategies that lead players to complete a game [62]. Other methods for segmenting players, notably funnel analysis [15], can be adapted for temporal bins but do not involve any analysis of gameplay - only player numbers surviving each progressive step in the funnel. Using the approach outlined here allows game developers and decision makers to observe (identify) and interact with variations in player behavior over time. This can serve e.g. as a benchmark for tracking disaffected players, compromised accounts, which types of players are at the highest risk for leaving the game, etc. Strong shifts in the data, either from players moving across clusters over time or player departure can indicate an issue that needs to be resolved or a disaffected group that is no longer satisfied with the game.

While the process for the research presented was manual, the methodology can be scaled up to be fully automated. This in turn allows for the diagram to be updated in concert with the specific needs of the client (e.g. running the analysis and re-building the diagram every 24 hours, or fully automating the system in real-time). As the populations in persistent or semi-persistent games change over time, building models that can handle new input data are important in game analytics [13,39]. Linking the model base code with a constantly evolving dataset, fed by an analytics client, and populating the Sankey diagram at selected frequencies, allows for monitoring the changes in the behavior of the player population

Future work will address the above potential for development, and also extend the current framework to other problem areas in game analytics and integrate multiple clustering algorithms for the purpose of evaluating both centroid-seeking behavior and extreme behaviors (e.g. using archetype analysis [12]). This will extend the usefulness of the current framework to locate extreme behavior clusters, such as gold farmers or cheaters, and evaluate their behavior across time. Furthermore, the *Glitch* dataset provides an opportunity for in-depth analysis about the health of the economy of the game, independent of player departure, and measuring financial indicators at the economy level (macroeconomic) [8,24,55] and the player level. Finally, on the visualization front, a next step is combining the auction house data with text mining analysis of collected forum data for a more holistic view of changes in player behavior; and rewriting the Sankey.js plugin code to enable acceptance of per player data as raw input would allow the construction of a system that enable drill-down analysis down to the user level, not just the level of individual clusters and time bins.

## 8. Conclusion

The goals of economic analysis of virtual economies are many, but fundamentally boil down to either gaining an understanding about how to monetize via virtual economies, how to make players enjoy in-game economies and take part in it, or investigate how it relates to the economics of the real world. The analysis presented here falls across these categories in that the focus is on player behavior and describing the economy in its own right, using explorative approaches rather than hypothesis-testing.

*Glitch* had a relatively short lifetime, and never reached the scale of *World of Warcraft, Lineage, Guild Wars, EverQuest II* or other million-plus subscriber MMOGs. The game represents the less well-known browser-based MMOs fairly well in terms of subscriber numbers, however, and its ultimate demise is also a well-known occurrence of the genre. Previous investigations of online game economies have not targeted the browser-based class of MMOGs, having instead mainly focused on major commercial MMOGs like *EverQuest II, World of Warcraft* and *Eve Online*.

In this paper, an explorative, data-driven analysis has been presented of how the *Glitch* players used the auction house – the nexus of the in-game economy in the game – across the lifetime of the game. Furthermore, we present a web-based interactive Sankey diagram, which shows the proportion of the different clusters across 14 time bins, documenting the flow of players between clusters as a function of time (diagram is available on: http://powerful-meadow-8588.herokuapp.com/).

The results of the analysis presented shows the pattern of decline in the game as player numbers dwindled, and documents the changes in the in-game economy from the period the game launched until it was taken off-line, covering almost three million data points.

The results also show how players change their use of the auction house across time, and document the migration of players into and out of the game in terms of how they used the auction house. Clusters of behavior are relatively consistent even during the decline in player numbers: The four high-level clusters (Casual, Moderate, Forum, Hardcore) survive a reduction in the player base from 4,632 unique monthly players to just 723 in the final month before *Glitch* closed. Retention rates fluctuate substantially across time, but apparently not as a direct effect of the *Glitch* closing down. Finally, the Hardcore cluster is the only one that consistently sees a large proportion of the constituent players migrating to the same behavioral cluster in succeeding months, unlike players in the Casual cluster, that exhibit high churn rates.